\documentstyle[twocolumn,aps,prl,epsf]{revtex}

\begin{document}

\title{{\bf Boundary conditions at a fluid -- solid interface}}

\author{{\sc Marek Cieplak$^{1,2}$, Joel Koplik$^{3}$ 
\& Jayanth R. Banavar$^{2}$ }}

\address{$^{1}$ Institute of Physics, Polish Academy of Sciences,
02-668 Warsaw, Poland} 

\address{$^{2}$ Department of Physics and Center for Materials Physics,
104 Davey Laboratory, The Pennsylvania State University, University
Park, Pennsylvania 16802}

\address{$^{3}$ Benjamin Levich Institute and Department of Physics,
City College of New York, NY 10031} 

\address{ $\;\;$}

\address{ $\;\;$}

\address{
\centering{
\medskip\em
{}~\\
\begin{minipage}{14cm}
We study the boundary conditions at a fluid-solid interface
using molecular dynamics simulations covering a broad
range of fluid-solid interactions and fluid densities,
and both simple and chain-molecule fluids. 
The slip length is shown to be independent of the type of flow,
but rather is related to the fluid organization near the solid, 
as governed by the fluid-solid molecular interactions.
{}~\\
{}~
\end{minipage}
}}

\maketitle

\newpage 

The principal theme of this paper is a study of the nature of
the boundary conditions (BC) of fluid flow past a solid surface,
a crucial ingredient in any continuum fluid mechanical calculation. 
The BC cannot be deduced from the continuum differential equations 
themselves, and it is often not easy to determine them experimentally.
While the normal component of the fluid velocity
must vanish at an impermeable wall for kinematic reasons,
the parallel component, when extrapolated toward
the wall, may match that of the wall only at some distance $\zeta$
away from it.
This phenomenon is known as slip and $\zeta$
is the slip length \cite{Kennard}.  Since the pioneering work
of Maxwell \cite{Maxwell}, it has been recognized that the
scale of $\zeta$ for a simple dilute gas
is set by the mean free path $\lambda$ of the fluid molecules,
with an O(1) proportionality constant for a thermalizing
wall.  
However, for a specularly reflecting wall, the proportionality
constant could become large and lead to large slip.  In the limit of low fluid
density, $\rho$, $\lambda$ becomes large suggesting
that $\zeta$ would be large as well. 
Furthermore, 
in this limit the continuum approximation need not hold, and it is hard 
to glean the nature of the BC without a detailed knowledge of the influence
of the fluid-solid interaction \cite{Sone}. This is indeed the situation for
micro-electro-mechanical systems \cite{Ho-Beskok} which operate
in the Knudsen regime, in which $\lambda$ can be larger than the system size.
Earlier molecular dynamics (MD) studies have
indicated substantial velocity slip  for repulsive walls and
deviations in hydrodynamic velocity profiles near the wall 
on lowering $\rho$ \cite{Willemsen,Robbins,Ebner,Troian}.

\vspace*{0.3cm}

In our MD simulations, we find that
the flow profile in the middle of a channel
does indeed correspond to that predicted by continuum theory, but we 
observe a range of behaviors near the walls.
Our work provides a molecular basis for the
large variability in the
amount of slip observed experimentally \cite{Lord}.
We find that $\zeta$ is an excellent descriptor of the boundary
conditions, independent of the channel width and the nature of the flow.
Even at high densities, significant slip is induced on weakening the
wall-fluid attraction.  In the low $\rho$ subcontinuum regime,
a large $\zeta$ is found in virtually all cases, except for
a chain molecule liquid and a strongly attractive wall.  These
two distinct classes of behavior lead to predictions amenable
to experimental test.  First, in a pressure driven flow, the speed
with which fluid is transported shows a maximum as $\rho$ is
varied for all the large slip situations and none when the slip
length remains small. Second, this maximum speed should
scale linearly with the channel width for the large-slip case,
but should scale quadratically (as in a dense fluid)
for the small slip situation.  Finally, our results indicate
a strong link between the degree of slip at low $\rho$
and the nature of fluid organization in the vicinity of
the channel walls.  In addition, we find analogous behavior in 
calculations of {\em thermal} slip, which arises when
the two bounding walls of a channel are held at different 
temperatures \cite{Tenenbaum}.

\vspace*{0.3cm}

In order to elucidate the BC as one lowers $\rho$ from the
dense liquid to the dilute gas regime, we
consider a channel geometry  and adopt the 
parametrization of Thompson and
Robbins \cite{Robbins}. 
There are two parallel walls 
in the $x-y$ plane and $N$ fluid atoms between the walls. 
Periodic BC are imposed along the $x$ and $y$ directions.
In the monatomic case, 
two fluid atoms separated by a distance $r$ interact with the Lennard-Jones
potential  $V_{LJ}(r)= 4 \epsilon
\left[ (\frac{r}{\sigma})^{-12} \;-\;
(\frac{r}{\sigma})^{-6} \right] \;$, where $\sigma$ is the size
of the repulsive core. As in Ref. \cite{Robbins}, the potential
is truncated at $2.2 \sigma$ (and shifted). 
The MD studies are performed at a temperature
$T=1.1 \epsilon/k_B$, which is just above the liquid-gas coexistence
region for this model. The characteristic MD time unit 
$\tau_0= \sqrt{m\sigma ^2/\epsilon}$, where $m$ is the atomic mass, 
is O(1 ps) for typical fluids.
The walls are constructed from two [001] planes of an fcc lattice
(96 atoms at each wall, with lattice constant $0.85 \sigma$),
with wall atoms tethered to fixed
lattice site by a harmonic spring with a large spring constant.
For the narrowest channel studied here, the fluid atoms
were confined to a volume $13.6 \sigma \times 5.1 \sigma$
in the $x-y$ plane and $L_0 = 12.75 \sigma$ between the inner faces of 
the walls.
The wall-fluid interactions are modelled by a distinct Lennard-Jones
potential $V_{wf}(r)= 16 \epsilon
\left[ (\frac{r}{\sigma})^{-12} \;-\; A \;
(\frac{r}{\sigma})^{-6} \right] \;$. The parameter $A$ determines the
properties of the wall and varies between 1 and 0, corresponding to
attractive and repulsive walls, respectively. 
Our earlier studies \cite{Physica} of collisions between fluid 
molecules and this wall have
quantitatively confirmed that $A$=1 corresponds to a
purely thermal wall, and $A$=0 to a purely specular one.

\vspace*{0.3cm}

The chain molecules comprising the polymeric fluid
are composed of $n=10$ Lennard-Jones atoms each.
The consecutive atoms along the chain are tethered by the FENE
potential \cite{Kroger}, $V_{FENE}=-\kappa /2 \, log[1-(r/r_0)^2]$, 
where $\kappa =30\epsilon$ and $r_0=1.5\sigma$, which 
together with $V_{LJ}$ keeps these atoms
around $r=\sigma$ apart. This system is studied at 
$T=1.6\epsilon/k_B$, which is above the liquid-vapor critical point,
and the molecular diameter is small compared to the channel width
(the radius of gyration of a molecule does not exceed $1.7 \sigma$).
For either fluid, the equations of motion were integrated using
a fifth-order predictor-corrector algorithm \cite{Allen}.
The spatial averaging was carried 
out in slabs of width $\sigma /4$ parallel to the $z$-axis. The high dilution
data points were time averaged over 3 million $\tau _0$ to reduce the noise.
The results given here were obtained with a 
Langevin thermostat \cite{Allen},  
but a Nose-Hoover thermostat \cite{Allen}  
was found to yield virtually identical data.
The $T$-dependent noise was applied in 
all directions during
equilibration and only in the $y$-direction when data were collected.
In the studies of thermal slip, the thermostat is
applied only to the wall atoms, where we
assign $T=1.1\epsilon/k_B$ to one
wall and $T=1.3 \epsilon/k_b$ to the other, and determine
the variation of $T$ across the channel.
\vspace*{0.3cm}

The density profiles for the thermal slip simulation of the monatomic
fluid are shown in Figure~1, for the attractive wall case ($A$=1).
One can identify several layers near the wall, but at the center
of the channel the profile becomes smooth. The value of density
in the center, $\rho$, is used in the remaining plots as a measure of  
the fluid density. (A layered density profile is also characteristic
of both monatomic and polymeric fluids without an imposed temperature
gradient \cite{Physica}). 
The density profile is asymmetric,
as illustrated by the behavior of 
the density difference in the second layers
between the cold and hot sides, $\Delta \rho _2 =\rho _{2c} - \rho _{2h}$,
shown in the inset of Figure 1.

\vspace*{0.3cm}

For both kinds of fluids, with and without
a temperature gradient, the density in the
first layer is largely $\rho$-independent, as shown in the top two
panels of Figure 2 for the uniform-$T$ case.
There is, however, a noticeable difference in the behavior of the
second layer. In the monatomic case, the
density in the second layer undergoes a steady buildup from a negligible 
value, when $\rho$=0.004$\sigma ^{-3}$
(corresponding to $N$=100, where 96 atoms coat the walls
and the remaining 4 move essentially ballistically within the
channel), to a saturation value  
when $\rho$ becomes large.
In the polymeric case, on the other hand, the second layer is almost
fully constructed even at  $\rho$=0.002$\sigma ^{-3}$ and then its density
varies only in a minor way. Thus the effective boundaries for the
monatomic fluid undergo rapid transformation as a function of $\rho$,
in contrast to the behavior of the chain molecule fluid. 
\vspace*{0.3cm}

\vspace*{0.3cm}

The top and bottom panels of Figure 3 describe viscous
and thermal slip, respectively, for the monatomic fluid. The velocity
profile of the top panel was obtained for Couette flow.
The central region has the expected
hydrodynamic behavior -- a linear velocity profile with constant shear stress
for Couette flow, and a parabolic velocity profile with a linear shear 
for gravity-driven flow (not shown).
While the temperature profiles in the thermal gradient case
are remarkably similar to the velocity profiles in Couette flow,
the dependence of the thermal slip length 
(defined analogously to the velocity slip length)
on $\rho$ is
non-monotonic in a way that reflects
the behavior of $\Delta \rho _2$.   As seen in the middle panel of 
Figure~4 for $A$=1, the slip is a minimum 
at $\rho \approx 0.2 \sigma ^{-3}$, just where $\Delta \rho _2$ 
is maximal (see inset of Figure 1).
Generally, the profiles predicted by continuum theory
are seen either to persist to a point close to the wall (no slip),
or else to terminate sooner, with a rapid change in profile around the 
second layer. It is
the latter situation which is associated with slip, and it arises
at very low $\rho$ in most cases.
Recently, Travis and Gubbins \cite{Travis} have performed
MD studies of velocity profiles for $A=\frac{1}{4}$ in
gravity driven flows in a different parameter range:  liquid densities, 
much narrower wall separation O(4$\sigma$), and
with an acceleration an order of magnitude larger than used here. 
Presumably due to these differences, 
these authors obtained triple peaked velocity profiles with
large contributions to the flow from the first layers. 

\vspace*{0.3cm}

The precise definition of the slip length $\zeta$ is given in terms of
the position, $z_0$, at which the extrapolated
fluid velocity  reaches that of the wall (non-zero in the Couette case), 
as measured relative
to the inner layer of the wall atoms. $\zeta$ is taken to be positive for
$z_0$ outside of the channel and negative otherwise.
Figure~4 summarizes our results on viscous and thermal $\zeta$
for monatomic and chain molecules.  Viscous slip was studied in
Couette and gravity driven flows and at various channel widths (two shown).
We find a large variation in boundary conditions on changing the 
material parameters, in agreement with experimental observations \cite{Lord}.
However, the value of $\zeta$ is found to be {\em transferable} in that it
neither depends on the kind of flow nor on the channel width.
For a range of high $\rho$ values for the monatomic fluid with wall
interaction $A$=1, $\zeta$   
is roughly equal to the negative of
the distance between the wall and the second layer.
At low $\rho$, however, there is a rapid $\sim 1/\rho$ growth
and a large slip.

\vspace*{0.3cm}

The slip length is a sensitive function of the properties of the
wall. When $A$=0, there is large slip even at
high fluid densities, which        
is qualitatively consistent
with the scenario of specular collisions with the wall.
It is also reminiscent of the large slip
found in MD studies of a nonwetting droplet on a solid surface \cite{Barrat}
However, for intermediate values of $A$
there is a complicated crossover behavior, as illustrated by the
$\zeta$ -- $\rho$ curve for $A=\frac{1}{4}$ and $\frac{3}{8}$.
In  Ref.~\cite{Troian}, a universal relationship between
$\zeta$ and shear stress was found in the dense fluid
regime on varying $V_{wf}$. Our large $\rho$ data on $\zeta$
can similarly be collapsed by rescaling but our control of $V_{wf}$
is through the parameter $A$ as opposed to the amplitude of the full
potential. The latter situation \cite{Troian} would lead to a vanishing
of the channel confinement in the zero amplitude limit.

\vspace*{0.3cm}

The right-hand panel of Figure 4 shows that the  
variation of the viscous slip length $\zeta$ with $\rho$ for chain molecules
is more sensitive to $A$ than for monatomic fluids.
For $A=1$, $\zeta$  becomes 
most negative as $\rho$ tends to zero, while
for $A=\frac{1}{4}$, $\zeta$ is large and positive even at high $\rho$ 
(20.6$\sigma$ for $\rho=0.79\sigma ^{-3}$).
This result is qualitatively consistent with the MD results \cite{Stevens}
on hexadecane in the vicinity of a solid surface, and with experiments on thin 
polymeric film drainage \cite{Horn}.

\vspace*{0.3cm}


For monatomic fluids,
the fluid velocity in the middle of the channel in a gravity
driven flow exhibits a maximum as a
function of $\rho$ \cite{Physica}.  In contrast, the
bottom panel of Figure 2 shows that, for chain molecules and $A$=1, the
dependence of $v_{max}$ on $\rho$ is monotonic even though
the viscosity at low $\rho$ is substantially similar
for the $n=1$ and $n=10$ fluids (data not shown).
The large $\rho$ behavior of  $v_{max}$ is dictated by the viscous
effects associated with a dense fluid.
The increase in $v_{max}$ with $\rho$ at low $\rho$ for the monatomic case
originates in the
active reconstruction of the second layer.
In contrast, for the chain molecule fluid,
the organization near the walls remains largely
unchanged (the chain molecules of the first two layers are stuck
to the wall) so that viscous effects
dominate the behavior of $v_{max}$  even at small densities.

\vspace*{0.3cm}


In summary, the two modes of behavior of $\zeta$ as $\rho$ decreases to
a small value, rapid growth or approach to a constant, are seen to be
associated with distinctly different dependences of
$v_{max}$ on $\rho$. They also appear to be linked to
the presence or absence of a density buildup in the
second layer. 
Strikingly, the apparent complexity and non-universality
of the BC is due to the microscale fluid
physics in the vicinity of the solid interface. 

\vspace*{0.3cm}

This work was supported by  KBN (Grant No. 2P03B-146-18),
NASA, an NSF MRSEC grant, and the Petroleum Research Fund 
administered by the American Chemical Society.

\begin{center}
{\bf Figure captions}
\end{center} 

\vskip 0.5cm

{\bf FIG. 1}
Density profiles for the monatomic fluid undergoing thermal slip
for walls with $A$=1. 
The tethering centers of the molecules comprising the inner wall
are at the edges of the figure, and
the values of density at the center of the channel, $\rho$, are indicated
next to the histograms. 
The peak which is the closest to the wall denotes
the first layer, and the second corresponds to the second maximum
plus the elevated bin next to it.
The inset shows the difference in the densities
of the second layers in the vicinities of the cold
and hot walls (the subscripts $2c$ and $2h$, respectively) as a function
of the central density.

\vskip 0.5cm

{\bf FIG. 2.} 
The top two panels show the density, $\rho _i \sigma ^3$, in the
first ($i$=1, square symbols) and second ($i=2$, circular symbols)
layers for the indicated values of $A$.
The top and middle panels are for the simple and polymeric fluids,
respectively. The bottom panel shows
the velocity of the polymeric fluid in the center of the channel vs.\ $\rho$,
in gravity driven flow with $g=0.01\epsilon/m\sigma$.
$L_0$ refers to a channel 
of standard size and $2L_0$ corresponds to 
channels with twice the width.
In all cases, the statistical errors are negligible. 
\vskip 0.5cm

{\bf FIG. 3}. 
{\bf Top:}
Velocity profiles in Couette flow for various $\rho$, in
the narrowest channel with attractive walls; the upper and 
lower walls move with velocities 
$\pm 0.1\sigma / \tau _0$, respectively.
The slip length for the high $\rho$ case is -1.7$\sigma$.
{\bf Bottom:}
The temperature profile 
in the thermal slip problem.
The open and closed circles correspond to
$\rho$=0.773 and 0.004
$\sigma ^{ -3}$ respectively. 
For $\rho$=0.193$\sigma ^{-3}$, only the central slope is indicated.

\vskip 0.5 cm

{\bf FIG. 4.}
$\zeta$ (measured in units of $\sigma$)
vs. $\rho$ for viscous and thermal slip phenomena, for various
values of $A$.
The horizontal lines labeled I and II indicate  
the locations of the first and second layers, respectively.
In all panels,
triangles correspond to a channel of doubled width, and the remaining
symbols are for the standard width. In the left and right panels,
the solid and open symbols
correspond to $\zeta$, as determined
from Couette and gravity driven
(acceleration 0.01$\epsilon /m\sigma$), flows respectively.

\vskip 0.5 cm

\newpage

\begin{figure}
\vspace*{0.5cm}
\epsfxsize=3.0in
\hspace*{-0.5cm}
\centerline{\epsffile{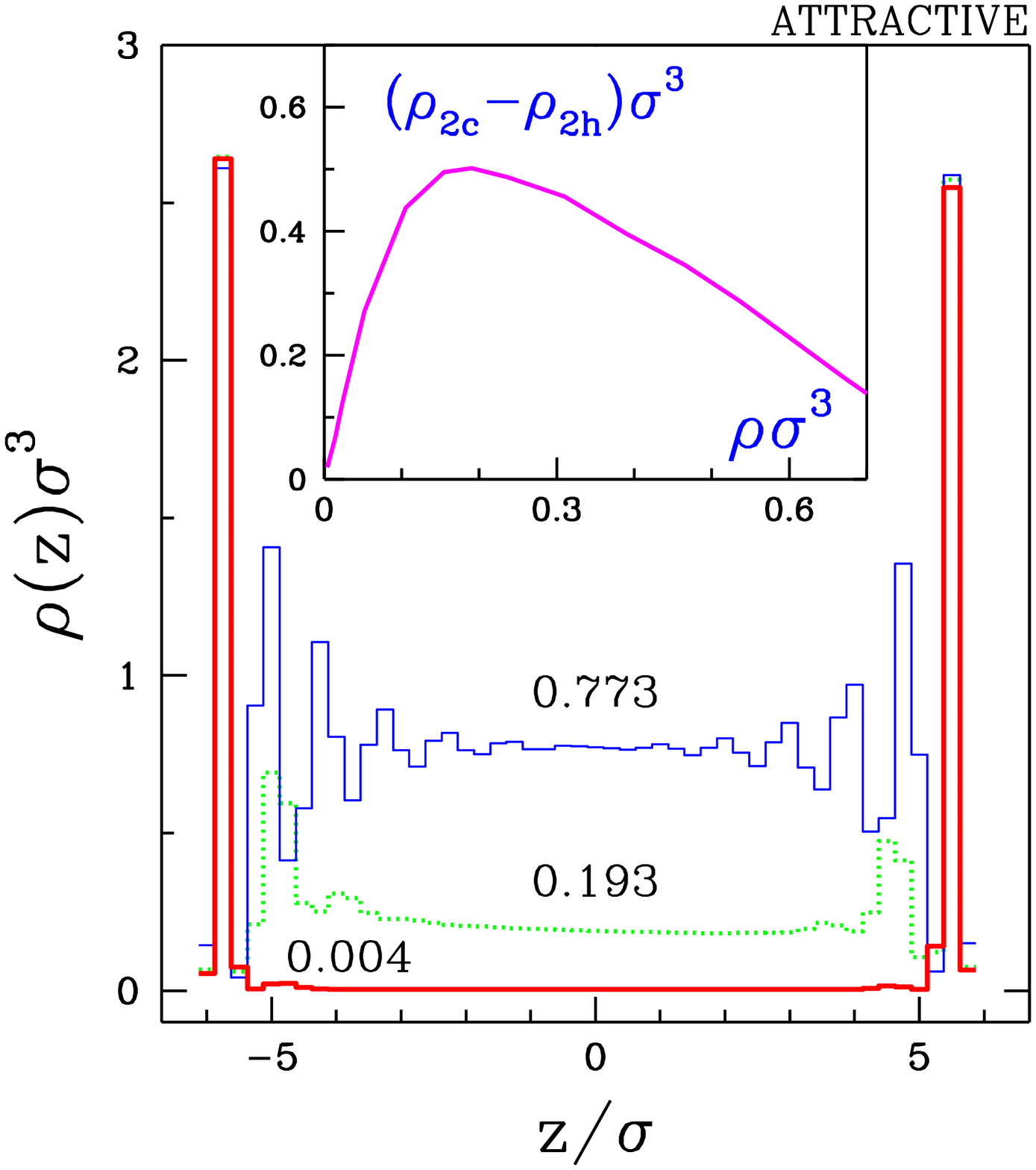}}
\vspace*{2.3cm}
\caption{ }
\end{figure}

\begin{figure}
\epsfxsize=3.8in
\centerline{\epsffile{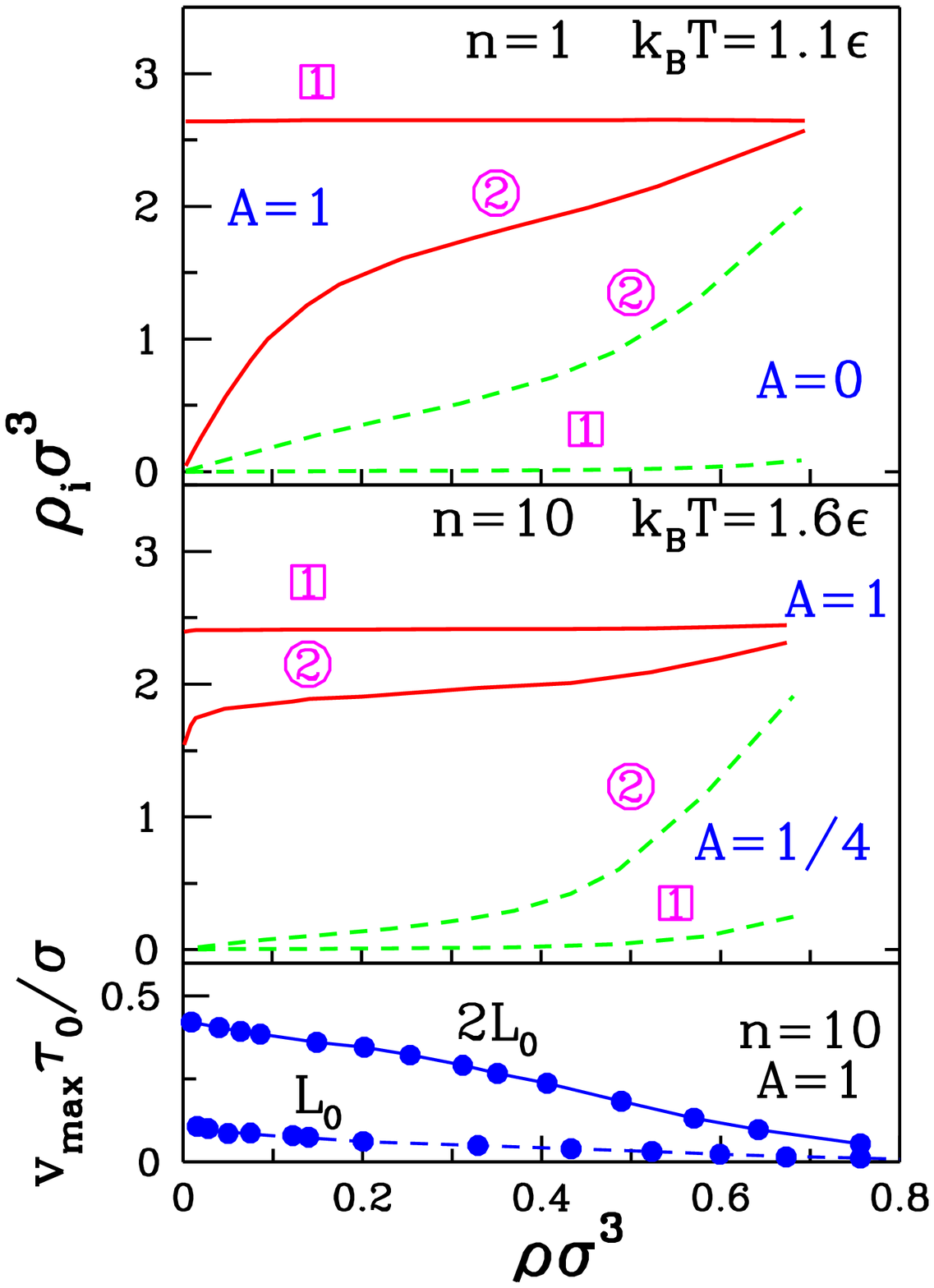}}
\vspace*{2.8cm}
\caption{ }
\end{figure}


\vspace*{1.5cm}
\begin{figure}
\epsfxsize=4.2in
\hspace*{2cm}
\centerline{\epsffile{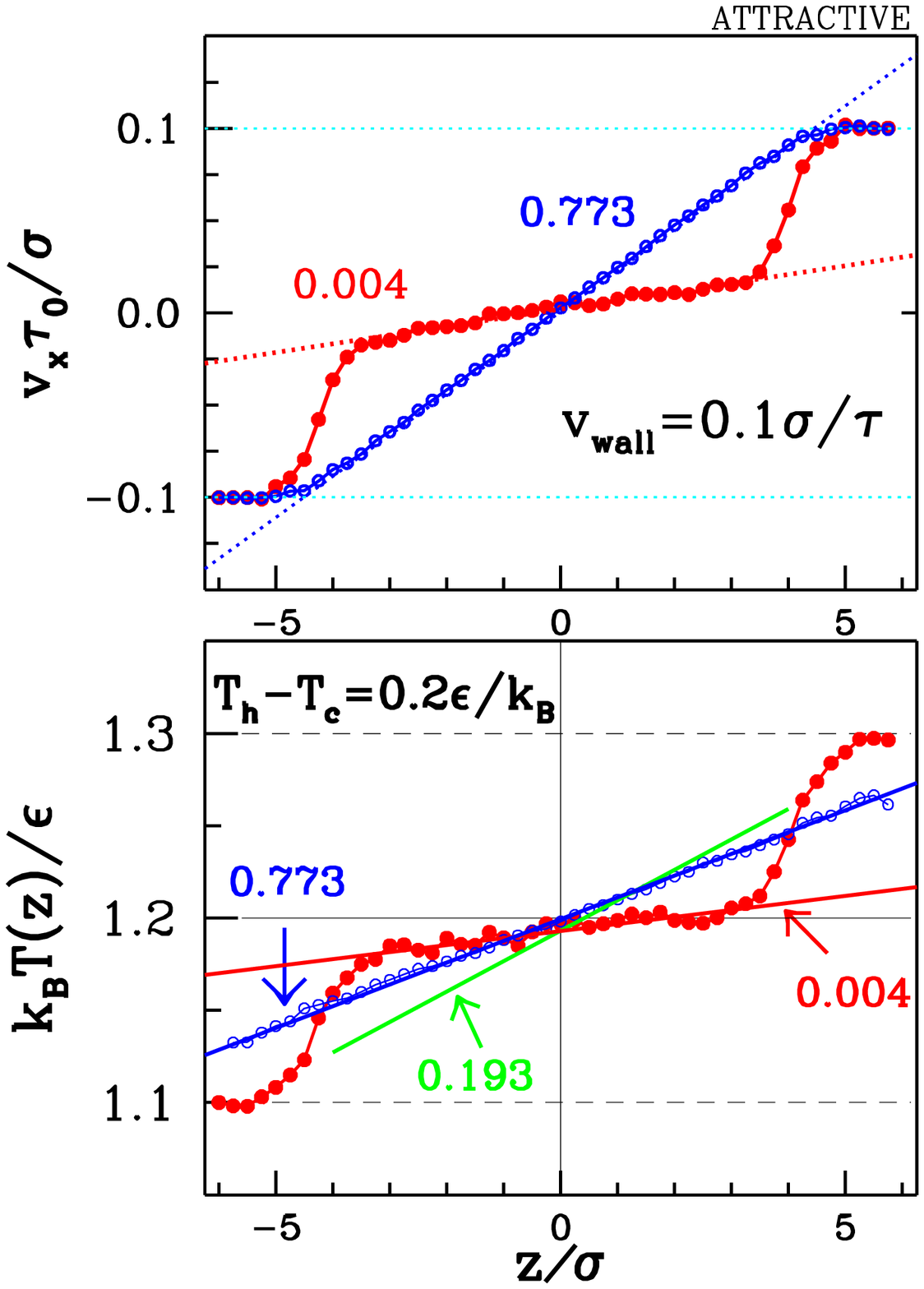}}
\vspace*{3.2cm} \hspace*{8cm}
\caption{ }
\end{figure}

\onecolumn

\begin{figure}
\epsfxsize=7.4in
\centerline{\epsffile{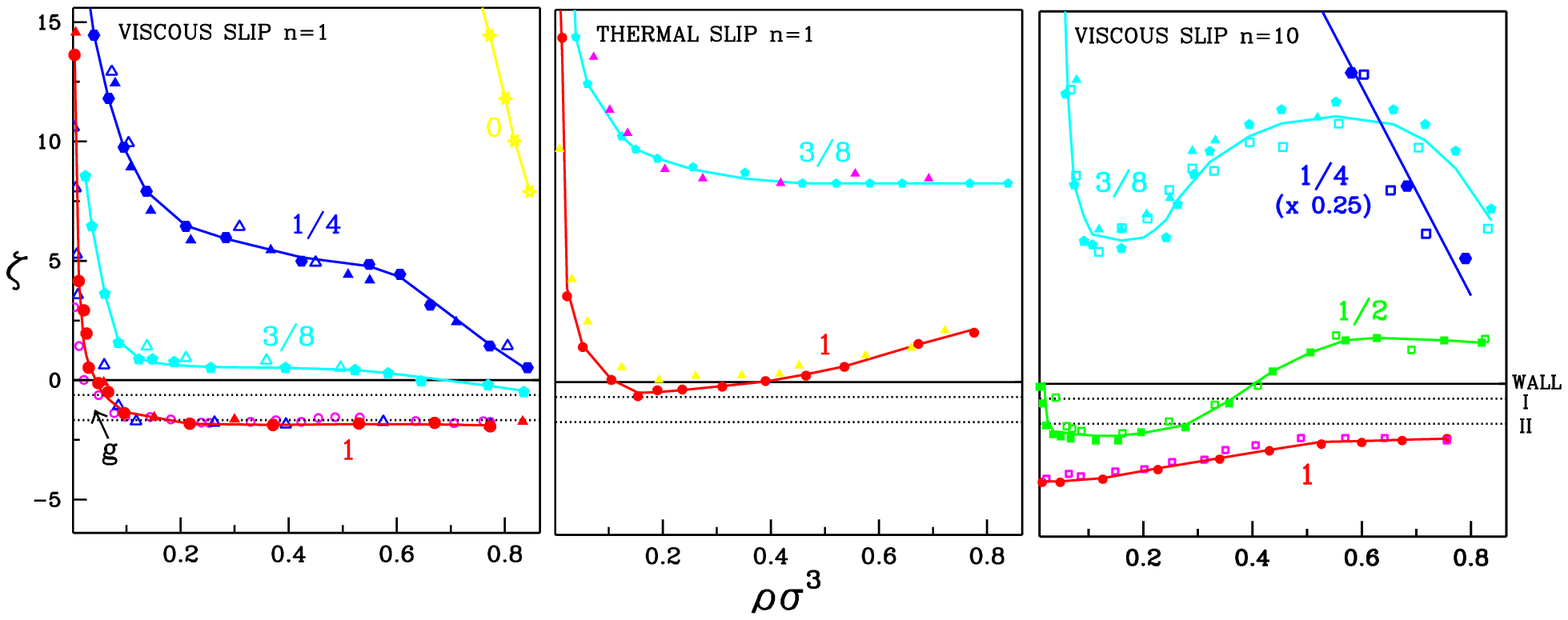}}
\vspace*{-15cm}
\caption{ }
\end{figure}

\end{document}